\documentclass[pra,showpacs,superscriptaddress,groupedaddress]{revtex4}
\usepackage{ae}
\usepackage[T1]{fontenc}
\usepackage[ansinew]{inputenc}
\usepackage{amsmath}
\usepackage{amssymb}
\usepackage[caption=false]{subfig}
\usepackage{multirow}
\usepackage{array}
\usepackage[]{graphicx}
\usepackage{wrapfig}
\usepackage{color}
\usepackage[colorlinks]{hyperref}
\usepackage{lscape}
\begin{document}
\title{Nature of Nonlocality in a triangle network based on EJM}
\author{Amit Kundu}
\email{amit8967@gmail.com}
\affiliation{Department of Applied Mathematics, University of Calcutta, Kolkata- 700009, India}
\author{Debasis Sarkar}
\email{dsarkar1x@gmail.com}
\affiliation{Department of Applied Mathematics, University of Calcutta, Kolkata- 700009, India}
\begin{abstract}
Defining nonlocality in a no-input closed quantum network scenario is a new area of interest nowadays. Gisin, in[Entropy 21, 325 (2019)], proposed a possible condition for non-tri-locality of the trivial no-input closed network scenario, triangle network, by introducing a new kind of joint measurement bases and a probability bound. In[npj Quantum Information (2020) 6:70] they found a shred of numerical evidence in support of Gisin's probability bound. Now based on that probability bound, we find the nature of the correlation in a triangle network scenario. We here observe how far the probability lies from that Gisin's bound with every possible combination of entangled and local pure states distributed from three independent quantum sources. Here we use the generalized Elegant Joint Measurements bases for each party and find that there is a dependency of non-locality on the entanglement of these joint measurement bases. We also check the probability bound for the polygon structure.
\end{abstract}
\date{\today}
	\maketitle
%	{Keywords: Quantum nonlocality; triangle network; elegant joint measurement; nonlocal resources.}
\section{Introduction}
Bell inequalities and their violation is a milestone in quantum theory. Violations of Bell inequalities state that quantum theory cannot be reproduced by any local hidden variable model\cite{bellpaper}. Such violations, referred to as quantum nonlocality, do not only provide insights into the foundations of quantum theory but also lead us to a broad scope of applications in quantum information science \cite{NDSVS}. In the usual  Bell scenario, two distant observers share a state from one single quantum source, who performs local and independent measurements on their respective particles. Bell demonstrated that the two observers can establish strong correlations which cannot be explained in any physical theory satisfying local model. Very recent long-waited loophole-free tests of quantum nonlocality were reported, providing the basis for the implementation of device-independent quantum information protocol\cite{GG7, GG8, GG9}.

Now, capturing nonlocality in the quantum network is not a very trivial one as the situation is qualitatively different from that of standard Bell nonlocal scenario. The basic difference between them is the number of sources distributing states between parties\cite{BNN}.  In \cite{GG11, GG12} they introduced bilocality in the simplest quantum network with two independent quantum sources sharing quantum states with three distant parties. They defined the bilocal correlation and reproduced a Bell-like inequality called bilocal inequality without input choices for the middle party, say, Bob, uses BSM( Bell State Measurement) on his two particles shared by two different sources. The violation of that inequality confirms that correlations to be non-bilocal in nature. Later, several works done on bilocal and its generalization n-local correlations \cite{GG13, my, tava, silva23}. Also, many works have been done on characterizing nonlocality in the no input scenario\cite{GG13, neural, causal, noinbi}. The simplest network structure in a closed loop network without input choices is a triangle network where three parties Alice, Bob and Charlie, in a triangle-shaped structure, share particles from three different sources between them [Fig: 1]. Finding the nature of correlation in this structure is really a hard task \cite{tr1, tri1, tri2, tri3}. In \cite{25y}, the authors stated that using three entangled states from three independent quantum sources and BSM as a measurement bases for each party the correlation given by the probability $p(a, b, c)$ is locally reproducible. They also introduced some joint measurement bases with a new kind of symmetry departed from the BSM called EJM(Elegant Joint Measurement). They found a probability bound using local correlation and got that the probability of getting the same outcome for each party(i.e., $p(a=b=c)$) using maximally entangled states and EJM, is somehow more than the bound. They also stated that only this probability will not conclude a correlation in triangle network non-trilocal but getting this high probability is an important indication. This check of nonlocality in the triangle network is characteristically different from the check of Fritz correlation \cite{GG13}, where they define nonlocality in the triangle network based on the Bell inequality between two parties. In Fritz correlation, any two classically correlated states and a maximally entangled state can produce non-trilocal correlations violating Bell inequality. Later, in \cite{causal, infla}, they tried to prove the concreteness of the probability bound of Gisin by some inflated technique and got a partially positive result. Recently in \cite{neural}, the authors numerically found evidence for some special cases in support of the results using neural network structure. In this work, considering the probability bound, like Fritz correlation, we try to find whether the combination of classically correlated states and entangled states coming from three different quantum sources $S_1$, $S_2$ and $S_3$ [see Fig: 1] can give such high probability of getting same outcomes and can possibly create a non-tri-local correlation using Elegant Joint Measurement as a measurement bases for each party. We use the generalized EJM and check the entanglement of the states based on a variable $\theta$ introduced in \cite{tavakoliBi}. We got a very surprising result about how the probability bound changes with variable entanglement of the measurement bases, which shows a very important connection between choices or the type of the joint measurement bases and the probability. There is a detailed comparison of the correlation probability coming from three different combinations of local and entangled quantum states with the EJM. We have also checked the probability bound for the polygonal structures. The whole paper is organized as follows: We first describe the motivation along with the generalized EJM in section II. Then we will provide the main result for the triangle scenario in Section III. There is some extension in the polygon scenario in section IV. This work ends with a discussion and practical implications.

\begin{figure}[h]
\centering
\includegraphics[width=8cm, height=8cm]{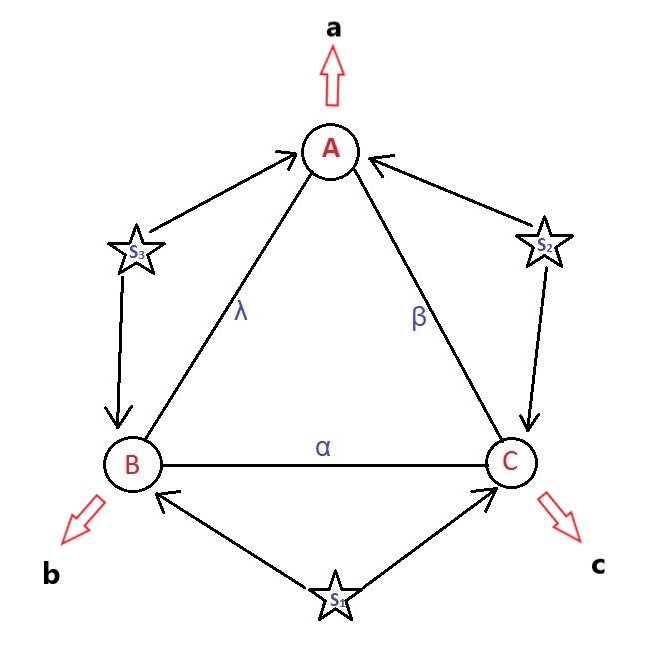}
\caption{The trivial closed quantum network with three parties A, B and C configured in a triangle manner. Here each party has no input choices and the outputs denoted by $a$, $b$ and $c$ from the fixed quantum measurement on the two particles coming from the three different sources, say, $S_1$, $S_2$ and $S_3$. On the other side, they share independent random variables $\alpha$, $\beta$ and $\gamma$ and output a function of random variables which they have access. The three random variables are local. Here, to find a quantum correlation $p(a,b,c)$ which can't be revealed by any three local scenario.}
\label{fig:triangle}
\end{figure} 

\section{Motivation and preliminaries}
In \cite{gisin17} the authors showed all pure states which are entangled, violate bilocal inequality and form nonlocal correlations in the bilocal network. For mixed states, it turned out that only one Bell nonlocal state can violate bilocal inequality. Later in \cite{andreoli,my}, they showed that for the more complex network, such as, Chain and Star networks, not all states coming from the quantum sources need to be Bell nonlocal to violate $n$-local inequality, where $n$ refers to the number of the sources.  By Bell nonlocal states we mean the states who violate Bell- CHSH inequality. In triangle configuration, the difference is, there are no input choices for each party and every party is connected and shared particles from two different sources as in the figure above. Each of the three parties is measuring fixed joint measurements on the two particles coming from two different independent quantum sources. The BSM (Bell State Measurement), we had so far used as a joint measurement setting, can not capture properly the quantumness in a triangle network \cite{25y}. The new kind of joint measurement settings with partially entangled bases showed some positive results about nonlocal behaviour in a triangle network \cite{25y}. They also found a probability bound of getting each party the same outcome using a local hidden variable model and showed that Elegant Joint Measurements with maximally entangled states can give the value of the probability very high from the bound. In \cite{neural}, they tried to prove the conjecture by some neural network technique and also found a positive supportive result. Using that bound we have tried to recognize the non-trilocal correlations more concretely with all pure entangled states shared by three independent sources and also with combinations of entangled states and product states shared by three sources. The joint measurement bases are playing an important role in these scenarios. We have investigated the correlation varying the entanglement of the Elegant Joint Measurement bases and we have noticed the changes in the probability for three different scenarios. We got interesting comparisons for three singlets states, two singlets with one product state and two product states with one singlet state. 

The whole structure of the new kind of joint measurement, different from BSM was introduced first in \cite{25y} with a two-qubit basis with four partially entangled eigenstates, all with the same degree of entanglement with some nice symmetries. They introduced four vertices of a regular tetrahedron inscribed in a Poincare sphere;
\begin{equation}
\begin{split}
&\vec{m_1}=(+1,+1,+1)\\
&\vec{m_2}=(+1,-1,-1)\\ 
&\vec{m_3}=(-1,+1,-1)\\
&\vec{m_4}=(-1,-1,+1)
\end{split}
\end{equation}
The states $|-\vec{m}_b\rangle$ directed to the antipodal direction. Specifically, this tetrahedron vertices can be written in cylindrical coordinates as $\vec{m}_b = \sqrt{3}(\sqrt{1-\eta^{2}_b}cos\phi_b,\sqrt{1-\eta^{2}_b}sin\phi_b,\eta_b)$ and can be defined as
$$|\vec{m}_b\rangle = \sqrt{\frac{1+\eta_b}{2}}e^{-i\phi_b/2}|0\rangle + \sqrt{\frac{1-\eta_b}{2}}e^{i\phi_b/2}|1\rangle$$
Note that, $m_b = \langle\vec{m}_b|\vec{\sigma}|\vec{m}_b\rangle$ (where $\vec{\sigma}$ be the vector formed by the three Pauli matrices). The newly introduced two-qubit bases constructed on anti-parallel spins are:
\begin{equation}
|\Phi_b\rangle = \frac{\sqrt{3}+1}{2\sqrt{2}}|\vec{m}_b, -\vec{m}_b\rangle + \frac{\sqrt{3}-1}{2\sqrt{2}}|-\vec{m}_b, \vec{m}_b\rangle
\end{equation}
The properties of the elegant measurement bases are:
$$\langle\Phi_b|\vec{\sigma}\otimes\mathbb{I}|\Phi_b\rangle = \frac{1}{2}|\vec{m}_b\rangle$$
and $$\langle\Phi_b|\mathbb{I}\otimes\vec{\sigma}|\Phi_b\rangle = -\frac{1}{2}|\vec{m}_b\rangle$$
Now in \cite{tavakoliBi}, generalized elegant joint measurement was introduced, based on a variable $\theta \in [0,\frac{\pi}{2}]$, that changes the entanglement of the bases from EJM to BSM. The generalized EJM bases, with the above properties, are
\begin{equation}\label{equ:EJM}
|\Phi^{\theta}_b\rangle = \frac{\sqrt{3}+e^{i\theta}}{2\sqrt{2}}|\vec{m}_b, -\vec{m}_b\rangle + \frac{\sqrt{3}-e^{i\theta}}{2\sqrt{2}}|-\vec{m}_b, \vec{m}_b\rangle
\end{equation}
It should be noted that for $\theta$ = $0$, the states are EJM introduced in \cite{25y}(the largest local tetrahedron in the family, of radius $\frac{\sqrt{3}}{2}$), while for $\theta = \frac{\pi}{2}$, the states are the usual BSM(Bell state Measurement, the smallest local tetrahedron, of radius zero) upto local unitaries. By varying $\theta$, the states can continuously jump between EJM and BSM. Now we can see the change of the entanglement of the bases, using von Neumann entropy of the reduced density operators, based on the variable $\theta$ (see FIG. 2):
\begin{figure}[h]
\centering
\includegraphics[width=9cm, height=7cm]{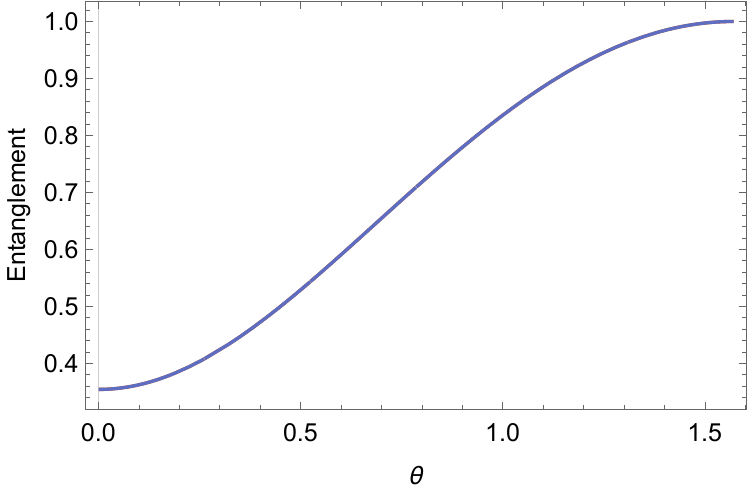}
\caption{From $\theta$=0 to $\theta$=$\frac{\pi}{2}$, where all the states jump from partially entanglement to maximally entanglement states.}
\end{figure}

Based on that we are going to check how the probability bound of getting same outcome for three parties in a triangle scenario, i.e., $p(a= b= c)$ changes with the changing of entangled bases.

\section{main results for triangle network based on EJM}
In \cite{tr2,25y}, considering three independent singlets from three independent sources, Gisin et. al. showed that if Alice, Bob and Charlie each perform the EJM on their two independent qubits the resulting correlation $p_{tr}(a,b,c)$, where $a,b,c = 1,2,3,4$ is fully characterized by three numbers corresponding to the cases $a = b = c,$ $a = b \neq c$ and $a\neq b\neq c$. It leads to the equations:
\begin{equation}
\begin{split}
&p_{tr}(a = k, b = k, c = k) = \frac{25}{256}~~ \text{for} \,k = 1,2,3,4\\
&p_{tr}(a = k, b = k, c = m) = \frac{1}{256}~~ \text{for} \,k \neq m\\
&p_{tr}(a = k, b = n, c = m) = \frac{5}{256}~~ \text{for} \,k \neq m\neq n\neq k
\end{split}
\end{equation}

For the normalization: $4\cdot\frac{25}{256} + 36\cdot\frac{1}{256} + 24\cdot\frac{5}{256} = 1$.
Here the global state considered is
$$|\Psi\rangle_{ABC}= \frac{1}{2\sqrt{2}}(|01\rangle-|10\rangle)_{AB}\otimes(|01\rangle-|10\rangle)_{BC}\otimes(|01\rangle-|10\rangle)_{CA};$$ So,
$$|\Psi\rangle_{AABBCC}=\frac{1}{2\sqrt{2}}(|101010\rangle - |001011\rangle - |101100\rangle + |001101\rangle - |110010\rangle + |010011\rangle + |110100\rangle - |010101\rangle).$$
Very simple steps show that $p_{tr}(a = b) = \frac{7}{16}$, $p_{tr}(a =k|b = c = k) = \frac{25}{28}$, and $p_{tr}(a = b = c)= \frac{25}{64}\approx 0.39$.
Now to answer the question, whether $p_{tr}(a, b, c)$ is tri-local or not, i.e., it obeys $$p_{tr}(a,b,c) = \int \int \int d\alpha  d\beta  d\gamma \mu(\alpha)\mu(\beta)\mu(\gamma)P_{A}(a|\beta\gamma)P_{B}(b|\gamma\alpha)P_{C}(c|\alpha\beta)$$ the above form or not, where $\mu(\alpha)$, $\mu(\beta)$ and $\mu(\gamma)$ are the densities for the classical shared randomness $\alpha$, $\beta$ and $\gamma$; they produced a tri-local model in \cite{25y} and showed the value of $p_{tr}(a=b=c)$ by local correlation and also showed that maximally entangled states and EJM as bases returns the $p_Q(a = b = c)$ much higher than the local value. In such a tri-local model of $p_{tr}(a,b,c)$ the correlation between Alice and Bob only due to their shared randomness $\gamma$. Similarly the other correlations between Bob-Charlie and Alice-Charlie are due to $\alpha$ and $\beta$. They considered a natural type of three-local model by making $\gamma = (\gamma_1, \gamma_2),$ where $\gamma_1 = (1,2,3,4)$ with equal probability and $\gamma_2 = 0, 1$ with $\text{prob}(\gamma_2 = 1) = q$ and similarly for $\alpha$ and $\beta$ also. Averaging the probabilities that $a = b = c$ over eight combinations of values of $\gamma_2$, $\beta_2$ and $\alpha_2$, 
the maximum probability bound they have found for $q = \frac{1}{2}$:
\begin{equation}\label{equ:nonl}
max\, p_{3local}(a = b = c) = \frac{61}{256}\approx 0.24
\end{equation}
\begin{figure}[h]
\centering
\includegraphics[width=9cm, height=7cm]{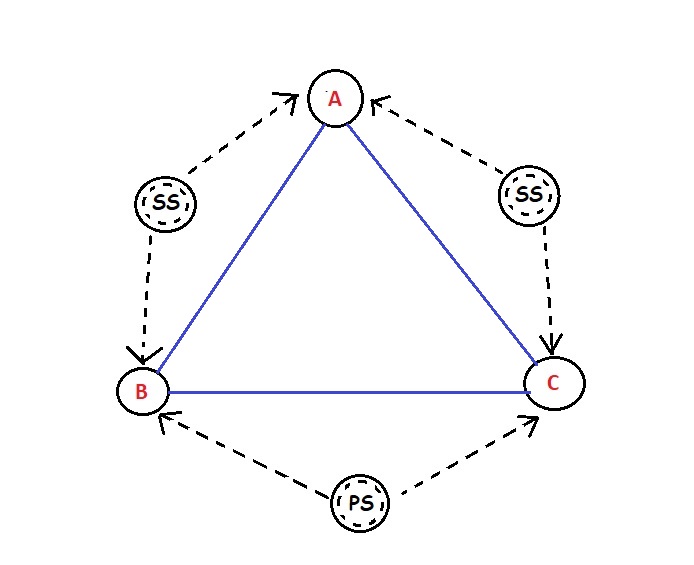}
\caption{The triangle network with one product state(PS) between Bob and Charlie and two Bell state(SS) between Alice, Bob and Alice, Charlie from three different independent sources.}
\label{fig:one}
\end{figure}
Now it is natural to ask that what if instead of using two-qubit maximally entangled states (i.e., Bell states), one uses partially entangled states? Further what we observe if all the sources are not sharing pure entangled states but one or two sources are sharing pure product states? Will there be still non-tri-local correlation between this tripartite scenario?
%Here we ask the question that if the non-tri-local correlation only comes from three independent singlets states or any less number of entangled states? What if all the sources are not sharing entangled states but one or two sources are sharing product states? Will there be still non-tri-local correlation between this tripartite scenario?\\

Consider first the case where we assume that the one source is sharing a product state between any two parties and the other two sources are sharing maximally entangled states. Then the global state [Fig:\ref{fig:one}] will be,
$$|\Psi\rangle_{ABC}= \frac{1}{2\sqrt{2}}(|01\rangle-|11\rangle)_{AB}\otimes(|01\rangle-|10\rangle)_{BC}\otimes(|01\rangle-|10\rangle)_{CA};$$ So,

$$|\Psi\rangle_{AABBCC}=\frac{1}{2\sqrt{2}}(|101010\rangle - |001011\rangle - |101100\rangle + |001101\rangle - |111010\rangle + |011011\rangle + |111100\rangle - |011101\rangle).$$
The statistics of the correlation between the parties are given by $p_Q(a,b,c) = |\langle\psi_a|\langle\psi_b|\langle\psi_c||\text{Bell state}\rangle|\text{Bell state}\rangle|\text{product}\rangle|^2$, where $a,b,c = 1,2,3,4$ from Elegant Joint Measurement (\ref{equ:EJM}). Without any difficulties we can show that the probability 
\begin{equation}
p_{ssp}(a = k, b = k, c = k) = \frac{41}{512} ~~ \text{for} \,k = 1,2,3,4,
\end{equation}
 
so $$p_{ssp}(a = b = c ) = \frac{41}{128}\approx 0.32$$ The $p_{ssp}$ stands for the correlation from the triangle scenario with two Bell states and one product state and it's all possible combination. And the result is significant in the sense that this probability is greater than the probability bound for tri-local correlation (\ref{equ:nonl}). According to this result, we can say that with EJM basis and one product state from one source out of three, the correlation can possibly be $non$-$tri-local$ in nature.

For the next step, we take now two product states and one Bell state between three parties. Alice-Bob and Bob-Charlie are sharing product states and suppose Alice-Charlie are sharing one Bell state. The three parties have fixed input, and performing Elegant Joint Measurement. With a simple calculation we get $p_{spp}(a = k, b = k, c = k) = \frac{17}{512} ~~ \text{for}~~ k = 1,2,3,4$. So $p_{spp}(a = b = c ) = \frac{68}{512}\approx 0.13$, which is less than the probability of (equ:\ref{equ:nonl}). So we find that with Bell-Product-Product states the correlation cannot reach the bound being tri-local correlation.

\begin{figure}[h]
\centering
\includegraphics[width=10cm, height=6cm]{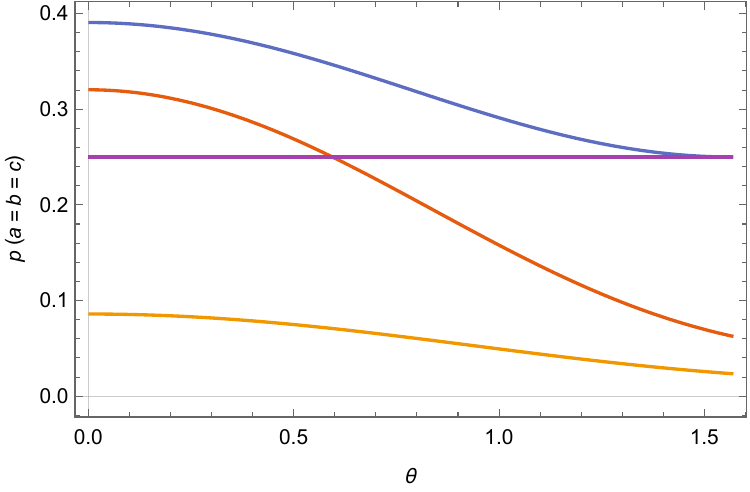}
\caption{The non-trilocality bound over the value of $\theta$. The violet line here represents the probability bound $\max p_{3local}(a = b = c) = \frac{61}{256}$. The Blue line represents the probability $p_{3local}(a = b = c) $ for three maximally entangled states scenario, merging with violet line at $\theta$ = $\pi/2$. The yellow line represents the probability for two product states and one maximally entangled states. While the orange line represents the probability for one product state and two entangled states. The crossing point of the orange line and the violet line is $(\frac{61}{256},\frac{65\pi}{314})$.}
\label{fig:theta}
\end{figure}

\begin{figure}[h]

	\begin{minipage}[c]{0.45\linewidth}
		\centering
		\includegraphics[width=7cm, height=5cm]{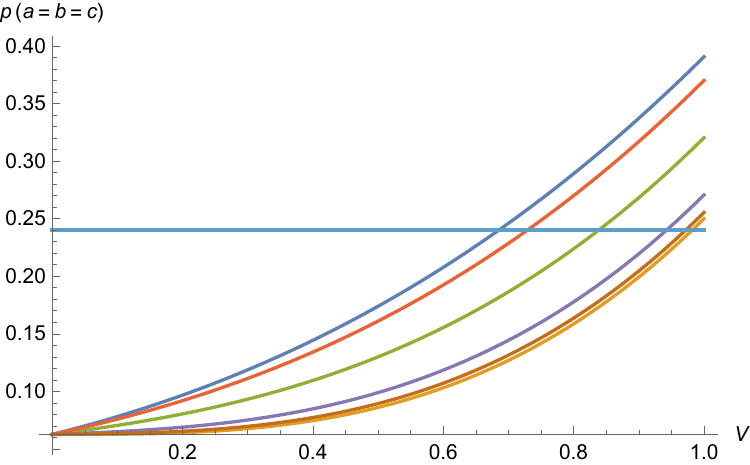}
		\caption{p(a=b=c) vs Noise V. Here V changes from 0 to 1. The blue line represents for $\theta$= $0$,Yellow line for $\theta$=$\frac{\pi}{2}$,Green line for $\theta$=$\frac{\pi}{4}$, red line for $\theta$=$\frac{\pi}{8}$, violet line for $\theta$=$\frac{3\pi}{8}$ and Brick red line is for $\theta$=$\frac{7\pi}{16}$.}
		\label{fig:noise}
	\end{minipage}
	\begin{minipage}[c]{0.45\linewidth}
		\centering
		\includegraphics[width=8cm, height=5cm]{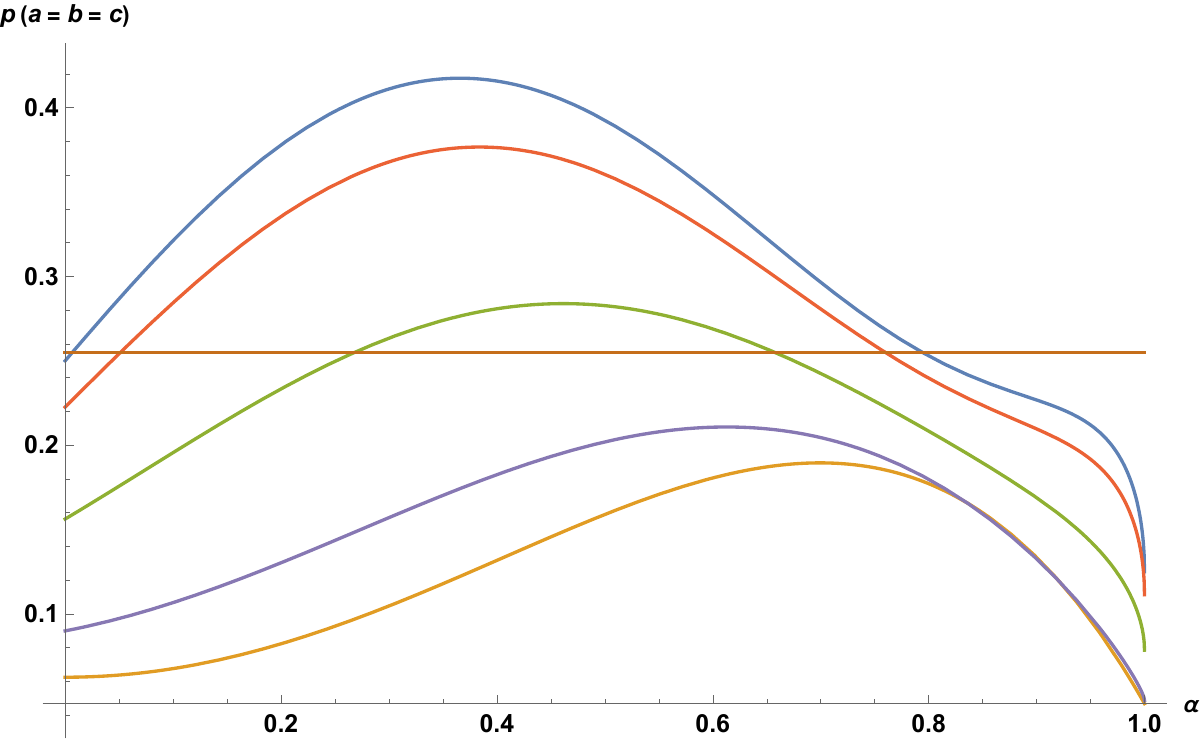}
		\caption{p(a=b=c) vs $\alpha$. Here blue line represents the changes for $\theta$=0, yellow line for $\theta$=$\frac{\pi}{2}$, green line for $\theta$=$\frac{\pi}{4}$, red line for $\theta$=$\frac{\pi}{8}$ and violet line for $\theta$=$\frac{3\pi}{8}$.}
		\label{fig:alpha}
	\end{minipage}
\end{figure}

As in the  (equ:\ref{equ:EJM}) we have a measurement choice parametrized by $\theta$ which can continuously shift from EJM to BSM and vice versa. Therefore there must be a $\theta$ dependency in this non-tri-locality. We find the probability $$p_{ssp}(a = b = c )$$ for different $\theta$ and have some exciting results for all three maximally entangled states, two entangled states one product states and two product one entangled state [fig:\ref{fig:theta}].
Now further if we share three states $|\psi\rangle =\alpha |01\rangle + \beta |10\rangle $ with the normalizations condition $\alpha^{2}+\beta^{2} = 1$ between three parties from three different sources, we get some significant outcomes [fig:\ref{fig:alpha}] on the probability $p(a=b=c)$. We notice that for $\theta$= 0 the probability crosses the no tri-local bound in the range of $\alpha \in(0.05-0.79)$. And [fig:\ref{fig:noise}] shows the dependency of the probability on the noise of Werner states $|\Psi\rangle = V|\psi^+\rangle\langle\psi^+|+\frac{1-V}{4}\mathbb{I}$ sharing between Alice Bob, Bob Charlie and Alice Charlie. In the range of $V\in(0.86, 1)$ the probability crosses the bound and shown non-tri-locality for EJM at $\theta=0$

\section{Correlation in the polygon type network with N=4}
\begin{figure}[h]
\centering
\includegraphics[width=10cm, height=7cm]{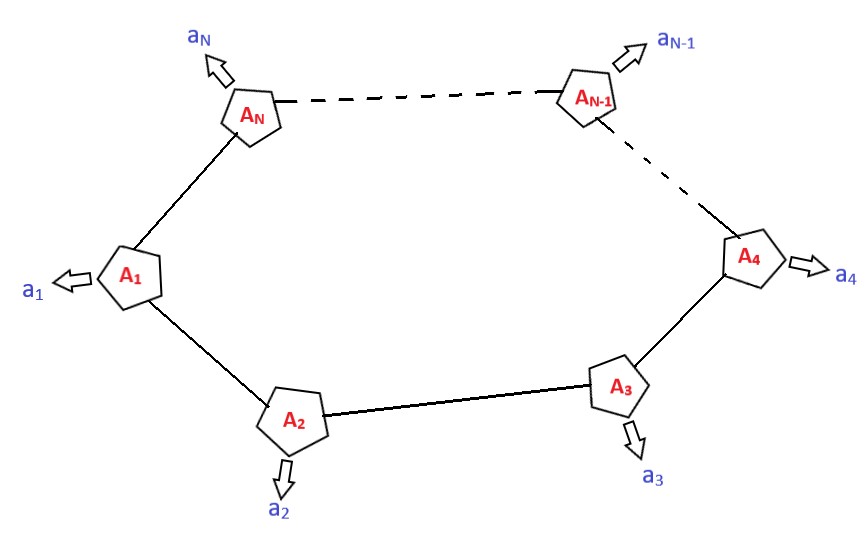}
\caption{The N-vertex polygon configuration for N parties. }
\label{fig:poly}
\end{figure}
The natural extension of closed network structure is a polygon with N vertices and N independent sources sharing states in between each pair parties connected by an edge [Fig.\ref{fig:poly}]. Here all the parties are performing EJM as same with the previous triangle scenario. The probability that all N parties in the polygon configuration get the same result was \cite{25y}, 
$$p_{polygon}(a_1 = a_2 = ... = a_N) = \frac{((-\sqrt{3}-1)^N + \sqrt{3}-1)^N)^2}{4^{2N - 1}}$$
Upto $N = 10$, the probability that all N outcomes are equal given that N-1 are equal was very large tending asymptotically to 0.933 and Gisin et. al. conjectured that $p_{polygon}(a_1,a_2,...,a_N)$ is not N-local for all $N \geq 3$.

\begin{figure}[h]
\centering
\includegraphics[width=9cm, height=7cm]{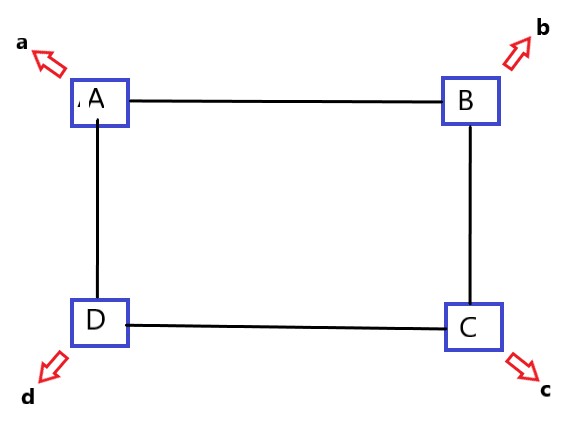}
\caption{The N = 4 configuration. }
\label{fig:square}
\end{figure}
Now consider a special case of a polygon with $N = 4$ [Fig:\ref{fig:square}] where all the parties are performing EJM with their states from the four independent sources. There is no probability bound for non N-locality in a polygon structure but there is a conjecture with the probability that all N outcomes are equal given that N-1 are equal is very large tending asymptotically to 0.933. We simply calculate this probability for $sssp(Bell state, Bell state, Bell state, Product)$, $sspp$ and $sppp$ scenario. We can not say which correlation shows non 4-locality but in future we need this value to fully characterize. The probability we find,
\begin{equation}\label{equ:4local1}
p_{sssp}(a_1 = ... = a_N|a_1 = ... = a_{N-1}) = \frac{5}{11}
\end{equation}
\begin{equation}\label{equ:4local2}
p_{sspp}(a_1 = ... = a_N|a_1 = ... = a_{N-1}) = \frac{416}{1024}
\end{equation}
\begin{equation}\label{equ:4local3}
p_{sppp}(a_1 = ... = a_N|a_1 = ... = a_{N-1}) = \frac{224}{512}
\end{equation}

\section{discussion}
So here we can make a comment that in a no-input closed network scenario, non-classicality is not only due to the sources distributing states to the parties but also depending on the nature of the measurement bases. Finding non-local correlation in the triangle network using Elegant Joint Measurement choices introduced by Gisin\cite{25y} instead of the existing Bell State Measurement confirms the point that the symmetry of the joint measurement bases is also important with the symmetry of the states from independent quantum sources, as the BSM cannot capture non-classicality in the trivial no-input closed network scenario. Gisin found a non-tri local correlation in the triangle scenario without inputs using EJM by defining a probability bound. We here checked the nature of the independent quantum sources, which helps the correlation to be non-trilocal between three parties in a triangle scenario. Is it necessary for the sources to produce only maximally entangled states to reach that high probability or one or two product states can create this high probability bound? The answer is no. It is not necessary. So when non-trilocality can be used as a resource to do any physical task the three parties can do it without knowing the full information of one source. For the measurement part, we also can say that the amount of entanglement plays an important role to produce non-tri locality in the triangle structure. So it is theoretically relevant and also practically very interesting to use the nonlocality in the trivial closed network scenario for some information processing protocol. The implementation of the EJM is thoroughly discussed in the \cite{tavakoliBi}. We can tune the entanglement in the states from the sources to get a non-trilocal correlation and also control the amount of entanglement of the states in the joint measurement bases. So the overall nature of the correlation is depending on two factors that we can control experimentally. Many works on the experimental realization of no input triangle scenario and its application have been done in recent times\cite{prx, brunner, post}. We can also ask about the nonlocality of the non-trivial closed network using EJM for further research. In this nontrivial no-input scenario, considering genuine nonlocality will also be a great future work\cite{trG}. Though in \cite{neural}, they have checked it with the noisy source states and noisy measurement bases, Implementing noise is also an interesting part of the closed network for four or more parties. Introducing FNN(Full Network Nonlocality) and source dependency in a closed network will be appreciable future work\cite{nn, l}. Also, non-trilocality can be checked with mixed states in future.

\section{acknowledgement}
A. K. acknowledges Guruprasad Kar for helpful discussion. A. K. also acknowledges support from CSIR, India and the author Debasis Sarkar acknowledges the work as part of Quest initiatives by DST India.
%\section{Conflict of interest} The authors declare no conflict of interest.
%%\newpage

\end{document}